\documentclass[aps,prd,preprint,superscriptaddress,amsmath,amssymb,showpacs]{revtex4-1}
\usepackage{dcolumn}
\usepackage{float}
\usepackage{physics}
\usepackage[colorlinks=true,allcolors=blue]{hyperref}
\usepackage{mathrsfs}
\usepackage{booktabs}
\usepackage{graphicx}
\usepackage{subcaption}

\begin{document}

\title{Thermal spectral functions of heavy and exotic Mesons from a hQCD at finite density}

\author{Zhou-Run Zhu }
\email{zhuzhourun@zknu.edu.cn}
\affiliation{School of Physics and Telecommunications Engineering, Zhoukou Normal University, Zhoukou 466001, China}

\author{Zhanyuan Gao }
\email{Zhanyuan@mails.ccnu.edu.cn}
\affiliation{Institute of Particle Physics and Key Laboratory of Quark and Lepton Physics (MOS), Central China Normal University,
Wuhan 430079, China}

\author{Zhaohui Feng }
\email{fengzh@mails.ccnu.edu.cn}
\affiliation{Institute of Particle Physics and Key Laboratory of Quark and Lepton Physics (MOS), Central China Normal University,
Wuhan 430079, China}

\author{Man-Li Tian}
\email{hbmutml@foxmail.com}
\affiliation{University Clinic, Zhoukou Normal University, Zhoukou 466001, China}

\author{Defu Hou }
\email{houdf@mail.ccnu.edu.cn}
\affiliation{Institute of Particle Physics and Key Laboratory of Quark and Lepton Physics (MOS), Central China Normal University,
Wuhan 430079, China}

\begin{abstract}
In this work, we study spectral functions of heavy and exotic mesons in a self-consistent Einstein-Maxwell-dilaton model at finite temperature and chemical potential, where the dilaton is allowed to back-react on the geometry. A non-quadratic dilaton profile is introduced to reproduce the nonlinear Regge trajectories of heavy-flavor mesons, and the corresponding spectral functions are computed for charmonium, bottomonium, tetraquark, and hybrid channels. In all cases, increasing either temperature or chemical potential suppresses and broadens the quasiparticle peaks, signaling sequential in-medium dissociation; for each radial family, the 2S-like state melts before the 1S-like state. The predicted melting temperatures obey the mass ordering $T_{\rm melt}^{\pi_1(1400)}<T_{\rm melt}^{J/\psi}<T_{\rm melt}^{Z_c(3900)}<T_{\rm melt}^{\Upsilon(1S)}$,
with $T_{\rm melt}\simeq 1.08,\ 2.96,\ 4.44$, and $5.93\,T_c$, respectively, so heavier states remain bound up to higher temperatures. These results provide quantitative, testable predictions for heavy-ion experiments probing tetraquark and hybrid mesons in hot and dense QCD matter.

\end{abstract}

\maketitle

\section{Introduction}\label{sec:01_intro}

Relativistic heavy-ion collisions at RHIC and the LHC make it possible to investigate
QCD matter under extreme conditions. In such collisions, the hot and dense medium is
expected to form the strongly coupled quark-gluon plasma (QGP) \cite{Matsui:1986dk,STAR:2019fge,Asakawa:2003re,Datta:2003ww,CMS:2018zza}. The properties
of the QGP can be probed through heavy quarkonium, whose suppression is regarded as a
signal of in-medium interactions \cite{Satz:2005hx}. Color screening in the QGP weakens the attraction
between the quark and the antiquark, causing the bound state to dissociate; correspondingly, its spectral function develops a suppressed peak with a broadened width.

Experimental data from Pb-Pb collisions at $\sqrt{s_{NN}} = 5.02$ TeV show that CMS
observed much stronger suppression of $\Upsilon(2S)$ and $\Upsilon(3S)$ than of
$\Upsilon(1S)$ \cite{CMS:2018zza}, a pattern known as sequential melting. The nuclear modification
factor of $J/\psi$ was also measured by ALICE \cite{ALICE:2019pid}. These results provide valuable
constraints on theoretical models of quarkonium dissociation. The internal structure
of a heavy meson further affects its response to the medium: a compact $q\bar{q}$
state such as $J/\psi$ behaves differently from a loosely bound molecular state such
as $Z_c$, or from a hybrid state such as $\pi_1$. The $Z_c$ states were discovered by
BESIII and LHCb \cite{BESIII:2013ris, LHCb:2021uow, BESIII:2015cld,LHCb:2015sqg}, and the $\pi_1(1600)$ hybrid was observed by COMPASS \cite{COMPASS:2009xrl}. However, the melting of such exotic states in a hot medium remains an open question.

Perturbative methods are not applicable in this regime, since QCD is strongly coupled at low energies, and lattice QCD suffers from the sign problem at finite chemical potential. Gauge/gravity duality offers a non-perturbative framework for exploring strongly coupled systems \cite{Maldacena:1997re,Witten:1998qj,Gubser:1998bc}. The melting of scalar glueballs and mesons has been
studied holographically \cite{Colangelo:2009ra,Miranda:2009uw}, and magnetic-field effects on quarkonium have also been explored \cite{Dudal:2014jfa,Braga:2018zlu,Zhao:2021ogc,Jena:2024cqs}. Dissociation at finite temperature and chemical potential was analyzed in \cite{Mamani:2013ssa,Braga:2016wkm,Braga:2017bml,Mamani:2022qnf,Toniato:2025gts}, and heavy-quarkonium masses and decay constants have been discussed as well \cite{Braga:2015jca,Braga:2015lck}. Other studies cover quasinormal modes \cite{Braga:2019yeh,Braga:2019xwl}, rotation effects \cite{Zhao:2023pne,Braga:2023fac,Zhu:2024uwu}, and anisotropy effects \cite{Chang:2024ksq}. Light-flavor hadron spectra have been analyzed within the dynamical holographic QCD model \cite{Chen:2022pgo}, the effect of $R^2$ corrections on spectral functions has been studied in Ref.\cite{Zhu:2024ddp}, and dissociation in a data-driven QCD model and in a gluon-condensate background has been explored in
\cite{Zhu:2026pqm,Zhao:2026bsc}.

Holographic QCD (hQCD) often employs the soft-wall model, in which a dilaton truncates AdS
space and reproduces linear Regge trajectories and confinement \cite{Karch:2006pv, Erlich:2005}. In this
approach, however, the dilaton usually does not back-react on the metric, making it challenging to describe the confinement-deconfinement transition and the meson spectrum simultaneously. Einstein--Maxwell--Dilaton (EMD) models provide such a framework, in which the Einstein, Maxwell, and dilaton equations are solved self-consistently \cite{Mahapatra:2020wym,Priyadarshinee:2021rch,Daripa:2024ksg,Cai:2012xh,He:2013qq,Dudal:2017max}; because the dilaton back-reacts on the geometry, the background and the meson spectrum are determined together. Recently, the authors of Ref.\cite{Toniato:2025gts} constructed a self-consistent dynamical holographic QCD model with a non-quadratic dilaton profile. However, a systematic comparison of the 1S and 2S spectral functions for mesons with different internal structures at finite density is still lacking. Here we adopt a different self-consistent EMD background that reproduces linear Regge trajectories and allows a flexible treatment of heavy and exotic states. Throughout this work, heavy mesons refer to charmonium and bottomonium, while the exotic sector includes the tetraquark \(Z_c\) and the light hybrid meson \(\pi_1\). The latter is a light-flavor state with explicit gluonic content, included here because its internal structure differs qualitatively from that of ordinary quarkonia.

In this work, we consider four representative vector-meson sectors: charmonium, bottomonium, tetraquark, and hybrid mesons. For each sector we focus on the two lowest radial excitations: $J/\psi$ and $\psi(2S)$ for charmonium, $\Upsilon(1S)$ and $\Upsilon(2S)$ for bottomonium, $Z_c(3900)$ and $Z_c(4200)$ for the tetraquark \cite{Yu:2024ljg}, and $\pi_1(1400)$ and $\pi_1(1600)$ for the hybrid meson \cite{ParticleDataGroup:2024cfk,Kopf:2020yoa}. Coupled-channel studies suggest that the latter two signatures might correspond to a single pole near 1560 MeV \cite{JPAC:2018zyd}; following Ref.~\cite{Toniato:2025gts}, we nevertheless treat them as the lowest two radial hybrid excitations. We examine how the temperature $T$ and the quark chemical potential $\mu$ affect the peak height, width, and position of the spectral functions, comparing the 1S and 2S states within each sector. Particular attention is paid to the differences between the 1S and 2S excitations and to the role of the chemical potential in modifying the dissociation behavior. The melting of ordinary quarkonia such as \(J/\psi\) and \(\Upsilon(1S)\) has been studied extensively, both on the lattice and in holographic models. In contrast, the response of exotic states such as the tetraquark \(Z_c\) and the hybrid meson \(\pi_1\) in a self-consistent EMD background at finite density remains largely unexplored. Because their binding involves multiquark correlations or gluonic flux, their dissociation pattern need not follow that of conventional quarkonia. The present work fills this gap by comparing all four sectors, charmonium, bottomonium, tetraquark, and hybrid meson, within a single self-consistent framework, and by providing testable predictions for the melting of \(Z_c\) and \(\pi_1\).

The paper is organized as follows. In Sec.~\ref{sec:02}, we briefly review the holographic EMD model. In Sec.~\ref{sec:03}, we study the spectral functions in holographic QCD. In Sec.~\ref{sec:04}, we give the conclusion and discussion.

\section{holographic EMD model}\label{sec:02}
We adopt the Einstein-Maxwell-Dilaton (EMD) \cite{Mahapatra:2020wym,Priyadarshinee:2021rch,Daripa:2024ksg,Cai:2012xh,He:2013qq,Dudal:2017max} framework to construct a self-consistent holographic model for heavy and exotic vector mesons. We briefly review the general structure of EMD systems. The five-dimensional EMD action is written as

\begin{equation}
\label{eqa112}
\begin{aligned}
S& =\frac{1}{16 \pi G_5} \int d^5 x \sqrt{-g}\left[R-\frac{f(\phi)}{4} F^2-\frac{1}{2} \partial_\mu \phi \partial^\mu \phi -V(\phi)\right], \\
\end{aligned}
\end{equation}
where \(f(\phi)\) is the gauge kinetic function, \(A_\mu\) denotes the Maxwell field, and \(V(\phi)\) represents the dilaton potential. These are determined self-consistently from the equations of motion. \(G_5\) denotes the five-dimensional Newton constant and we set it to be one in following calculations. \(\phi\) represents the dilaton field.

The metric ansatz is taken to be
\begin{eqnarray}
\label{eq18}
d s^2=\frac{L^2 e^{2 A(z)}}{z^2}\left[-g(z) d t^2+\frac{d z^2}{g(z)}+d \vec{x}^2\right],
\end{eqnarray}
where \(L\) is the AdS radius, and we take \(L=1\) for convenience.

With this ansatz, the background fields satisfy the following equations of motion and constraints
\begin{equation}
\begin{gathered}
\phi^{\prime \prime}+\phi^{\prime}\left(-\frac{3}{z}+\frac{g^{\prime}}{g}+3 A^{\prime}\right)-\frac{ e^{2 A}}{z^2 g} \frac{\partial V}{\partial \phi}+\frac{z^2 e^{-2 A} A_t^{\prime 2}}{2 g} \frac{\partial f}{\partial \phi}=0,
\end{gathered}
\end{equation}
\begin{equation}
A_t^{\prime \prime}+A_t^{\prime}\left(-\frac{1}{z}+\frac{f^{\prime}}{f}+A^{\prime}\right)=0,
\end{equation}
\begin{equation}
g^{\prime \prime}+g^{\prime}\left(-\frac{3}{z}+3 A^{\prime}\right)-e^{-2 A} A_t^{\prime 2} z^2 f=0,
\end{equation}

\begin{equation}
\begin{split}
A^{\prime \prime} +\frac{g^{\prime \prime}}{6 g}+A^{\prime}\left(-\frac{6}{z}+\frac{3 g^{\prime}}{2 g }\right)-\frac{1}{z}\left(-\frac{4}{z}+\frac{3 g^{\prime}}{2 g}\right) +3 A^{\prime}+\frac{ e^{2 A} V}{3 z^2 g}=0,
 \end{split}
\end{equation}

\begin{equation}
A^{\prime \prime}-A^{\prime}\left(-\frac{2}{z}+A^{\prime}\right)+\frac{\phi^{\prime 2}}{6}=0.
\end{equation}

One can obtain
\begin{equation}
\begin{aligned}
g(z) & =1+\frac{1}{\int_{0}^{z_h} y^{3}e^{-3A}\,dy}\left[-\int_{0}^{z} y^{3}e^{-3A}\,dy+\left(\frac{\mu}{\int_{0}^{z_h}\frac{y}{e^{A}f}\,dy}\right)^{2}\operatorname{det} \mathcal{G}\right],\\
\phi^{\prime}(z) & =\sqrt{6\left(A^{\prime 2}-A^{\prime \prime}-2 A^{\prime} / z\right)}, \\
V(z) & =-3 z^2 g e^{-2 A}\Big[A^{\prime \prime}+A^{\prime}\left(3 A^{\prime}-\frac{6}{z}+\frac{3 g^{\prime}}{2 g}\right) \\
& -\frac{1}{z}\left(-\frac{4}{z}+\frac{3 g^{\prime}}{2 g}\right)+\frac{g^{\prime \prime}}{6 g}\Big],
\end{aligned}
\end{equation}
where
\begin{equation}
\operatorname{det} \mathcal{G}=\left|\begin{array}{ll}
\int_{0}^{z_h} y^3 e^{-3A} dy & \int_{0}^{z_h} y^3 e^{-3A} dy \int_{0}^{y} \dfrac{x}{e^{A} f} dx \\
\int_{z_h}^{z} y^3 e^{-3A} dy & \int_{z_h}^{z} y^3 e^{-3A} dy \int_{0}^{y} \dfrac{x}{e^{A} f} dx
\end{array}\right|.
\end{equation}

The Hawking temperature is given by
\begin{equation}
\label{eqa1b}
\begin{aligned}
T =& \frac{1}{4\pi} \frac{z_h^3 e^{-3A(z_h)}}{\displaystyle\int_{0}^{z_h} y^3 e^{-3A(y)} dy}
\left[ 1 - \left( \frac{\mu}{\displaystyle\int_{0}^{z_h} \frac{y}{e^{A}f} dy} \right)^{2} \right. \\
&\left. \times \left( \int_{0}^{z_h} y^3 e^{-3A} dy \int_{0}^{z_h} \frac{x}{e^{A}f} dx
- \int_{0}^{z_h} y^3 e^{-3A} \int_{0}^{y} \frac{x}{e^{A}f} dx\,dy \right) \right],
\end{aligned}
\end{equation}
where $\mu$ denotes the baryon chemical potential.

The entropy is
\begin{equation}
s = \frac{L^3 e^{3A(z_h)}}{4G_5 z_h^3}.
\end{equation}

One can take the following form of wrap factor $A(z)$ \cite{Dudal:2017max}
\begin{eqnarray}
A(z)=-\frac{3}{4} \ln \left(a z^2+1\right)+\frac{1}{2} \ln \left(b z^3+1\right) -\frac{3}{4}  \ln \left(a z^4+1\right).
\end{eqnarray}

The gauge kinetic function $f(z)$ as
\begin{equation}\label{fff}
f(z)=e^{-A(z)-B(z)},
\end{equation}
where $B(z)=(\kappa z)^{2-\alpha} + Mz + \tanh\left(\frac{1}{Mz} -\frac{\kappa}{\sqrt{\Gamma}}\right)\Big]$ \cite{MartinContreras:2021bis}. We fix \(B(z)\) by fitting the mesonic mass spectra.

The two functions \(A(z)\) and \(B(z)\) play distinct roles. At zero chemical potential, the Hawking temperature (Eq.(\ref{eqa1b})) becomes
\begin{equation}
\label{eqa11}
\begin{aligned}
T =& \frac{1}{4\pi} \frac{z_h^3 e^{-3A(z_h)}}{\displaystyle\int_{0}^{z_h} y^3 e^{-3A(y)} dy},
\end{aligned}
\end{equation}
where only depends on the \(A(z)\), while is independent of \(B(z)\) and of \(f(z)\).

At \(T=\mu=0\), one has \(g(z)=1\), and the vector sector reduces to the action and fluctuation equation given by Ref.~\cite{Toniato:2025gts}. Starting from the EMD action (Eq.(\ref{eqa112}))
\begin{equation}
S=-\frac{1}{16\pi G_5}\int d^5x\sqrt{-g}\frac{f(\phi)}{4}F_{MN}F^{MN},
\end{equation}
and the corresponding equation of motion
\begin{equation}
\partial_M\left(\sqrt{-g}f(\phi)F^{MN}\right)=0,
\end{equation}
the zero-temperature metric becomes
\begin{equation}
ds^2=\frac{L^2e^{2A(z)}}{z^2}\left(-dt^2+dz^2+dx_idx^i\right).
\end{equation}

With the radial gauge \(A_z=0\) and the transverse Landau gauge \(\partial_\nu A^\nu=0\), the fluctuation equation reads
\begin{equation}
\frac{ze^{-A}}{Lf}\partial_z\left(\frac{Le^{A}f}{z}\partial_z A^\mu\right)+\Box A^\mu=0 .
\end{equation}

After a Fourier transform and the substitution \(A^\mu=e^{-H}\psi^\mu\) with
\begin{equation}
H=\frac12\ln\left(\frac{Le^{A}f}{z}\right),
\end{equation}
this turns into the Schr\"{o}dinger-like equation
\begin{equation}
-\partial_z^2\psi+U(z)\psi=m_n^2\psi,
\end{equation}
where
\begin{equation}
U(z)=\left(\partial_z\left[\frac12\ln\left(\frac{Le^{A}f}{z}\right)\right]\right)^2
+\partial_z^2\left[\frac12\ln\left(\frac{Le^{A}f}{z}\right)\right].
\end{equation}

Since \(f(z)=e^{-[A(z)+B(z)]}\), the combination that enters the potential is
\begin{equation}
e^{A(z)}f=e^{-B(z)}.
\end{equation}

Thus the mass spectrum at \(T=\mu=0\) depends only on \(B(z)\). This is why \(B(z)\) can be fixed by fitting the mesonic mass spectrum, while \(A(z)\) is fixed separately by the thermodynamic background. In this work,  we fix the parameters of \(A(z)\) by matching the confinement-deconfinement transition at \(T \simeq 270\,\mathrm{MeV}\) for \(\mu = 0\), giving \(a = 0.1232\) and \(b = 0.3625\).

The values of $B(z)$ parameters for the four meson states are in the following Table1 and reproduce the Regge trajectories. The parameters \(\kappa, M, \sqrt{\Gamma}, \alpha\) in \(B(z)\) are taken from Ref.~\cite{Toniato:2025gts}. Both linear and nonlinear Regge trajectories were fitted to the meson masses. While both provide good descriptions, the nonlinear fit is preferred because it reproduces the spectra with \(R^2=1\) and smaller deviations. We therefore adopt the nonlinear Regge parameters throughout this work. Reference~\cite{Toniato:2025gts} further shows that the mass spectra of \(J/\psi\), \(\Upsilon\), \(Z_c\), and \(\pi_1\) are reproduced with very small relative errors, indicating that the nonlinear Regge trajectories provide a reliable fit to the heavy and exotic meson spectra. Following Ref.~\cite{Toniato:2025gts}, the tetraquark \(Z_c\) is modeled as a hadronic molecule of \(J/\psi\) and \(\rho\), with averaged constituent mass \(\bar m_{\rm tetra}=0.283\,m_{J/\psi}+0.717\,m_{\rho}\). The hybrid meson \(\pi_1\) is modeled as a \(q\bar q g\) state with \(\bar m_{\rm hybrid}=0.497\,m_u+0.497\,m_d+0.006\,m_G\). These averaged masses fix the Regge parameters \(\alpha\) and \(\kappa\) in \(B(z)\). The internal structure thus enters only through \(\bar m\); no explicit multibody wave function is introduced.

\begin{table}[H]
\centering
\begin{tabular}{lcccc}
\hline
 & \(\kappa\)(GeV)  & \(M\)(GeV)  & \(\sqrt{\Gamma}\)(GeV)  & \(\alpha\) \\
\hline
Charmonium & 1.8   & 1.7   & 0.53  & 0.54  \\
Bottomonium & 9.9   & 2.74  & 1.92  & 0.863 \\
Tetraquark      & 1.75   & 1.44   & 0.3   & 0.539   \\
Hybrid meson    & 0.468 & 0.20  & 0.12  & 0.034 \\
\hline
\end{tabular}
\caption{Values of B(z) parameters for the four meson states, taken from Ref.~\cite{Toniato:2025gts}.}
\label{tab:parameters}
\end{table}

The spectral functions are computed for all four states at finite temperature and chemical potential. For the thermodynamic quantities (Figs.1-3), we use the hybrid meson \(\pi_1\) parameters (\(\kappa=0.468\) GeV, \(M=0.20\) GeV, \(\sqrt{\Gamma}=0.12\) GeV, and \(\alpha=0.034\)) to study the black hole thermodynamic. The qualitative conclusions for the black hole thermodynamics are unaffected by this choice.

\begin{figure}[H]
    \centering
    \begin{subfigure}[b]{0.48\textwidth}
        \centering
        \setlength{\abovecaptionskip}{0.1cm}
        \includegraphics[width=\linewidth]{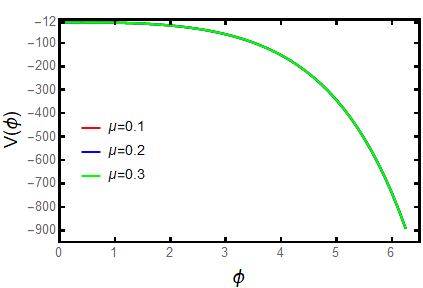}
        \label{fig:V1}
    \end{subfigure}
    \hfill
    \begin{subfigure}[b]{0.48\textwidth}
        \centering
        \setlength{\abovecaptionskip}{0.1cm}
        \includegraphics[width=\linewidth]{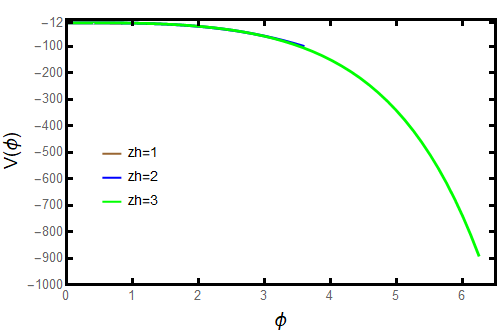}
        \label{fig:V2}
    \end{subfigure}
    \caption{Left panel: $V(\phi)$ as a function of $\phi(z)$ with different values of chemical potential for $z_h=3 GeV^{-1}$. Right panel: $V(\phi)$ as a function of $\phi(z)$ with different values of $z_h$ for $\mu=0.1 GeV$.}
    \label{fig1}
\end{figure}

A possible concern is that any dependence of \(V(\phi)\) on \(z_h\) or \(\mu\) would mean different actions for different \(z_h\) or \(\mu\). We examine how \(V(\phi)\) depends on the horizon radius \(z_h\) and the chemical potential \(\mu\) in Fig.~\ref{fig1}. If \(V(\phi)\) were sensitive to these parameters, the action and the metric solution would change accordingly. In Fig.~\ref{fig1}, the curves for different \(z_h\) and \(\mu\) overlap almost completely. This implies that \(V(\phi)\) is practically independent of \(z_h\) and \(\mu\), and the \(z_h\) or \(\mu\) dependence can be negligible.

\begin{figure}[H]
    \centering
      \setlength{\abovecaptionskip}{0.1cm}
    \includegraphics[width=10cm]{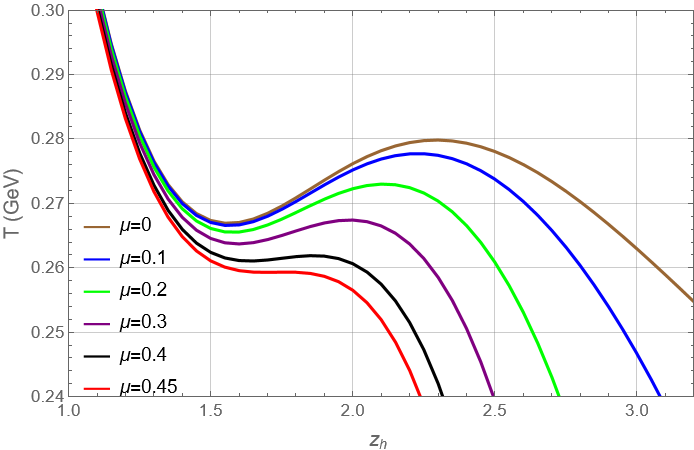}
    \caption{\label{fig2} The Hawking temperature \(T\) as a function of the horizon radius \(z_h\).}
\end{figure}

\begin{figure}[H]
    \centering
      \setlength{\abovecaptionskip}{0.1cm}
    \includegraphics[width=10cm]{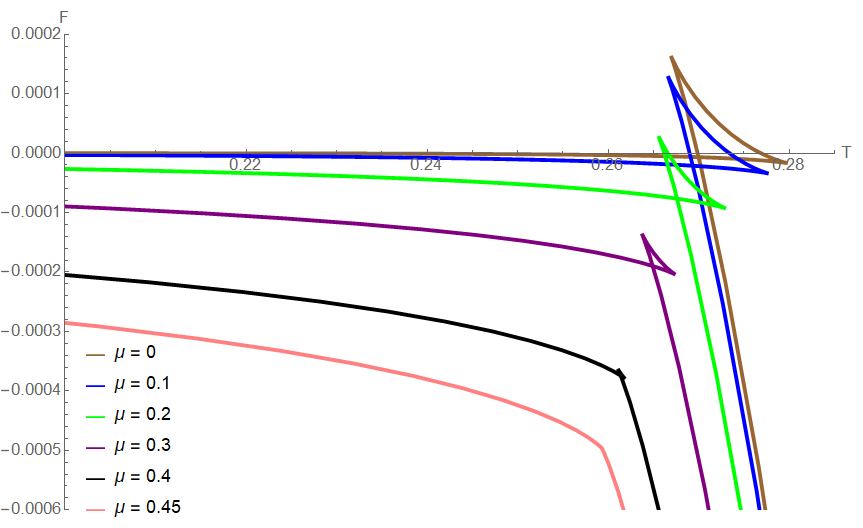}
    \caption{\label{fig3} The free energy \(F\) as a function of temperature \(T\).}
\end{figure}

The black hole free energy at fixed chemical potential is
\begin{equation}
\label{eqn1}
 F=\int^\infty_{z_h} s \frac{dT}{d z_h}d z_h.
\end{equation}

In Fig.~\ref{fig2} we plot \(T\) as a function of \(z_h\) for different values of \(\mu\). Three black hole branches can emerge: a large black hole at small \(z_h\), an unstable phase, and a small black hole at large \(z_h\). The large and small black holes are thermodynamically stable, while the intermediate unstable phase is not. For the large and small black holes, \(T\) decreases as \(z_h\) increases. For the intermediate unstable branch, \(T\) rises with \(z_h\). This branch has negative specific heat and negative squared sound speed. As \(\mu\) grows, the intermediate branch becomes smaller and the large and small branches approach each other. At a critical value \(\mu=\mu_c\), the intermediate branch disappears completely and the large and small branches merge into a single branch. For \(\mu>\mu_c\), only one branch remains, which becomes extremal at some horizon radius.

Fig.~\ref{fig3} shows the free energy \(F\) versus temperature \(T\) for several values of the chemical potential. For \(\mu<\mu_c\), the free energy takes a swallow-tail form, which signals a first-order transition. At high temperature the large black hole has lower free energy, whereas at low temperature the small black hole is favored. Consequently, a first-order transition between the two phases occurs as the temperature changes. The transition temperature decreases with increasing \(\mu\). The intermediate branch, which lies at the bottom of the swallowtail, always has higher free energy than the other two and is never thermodynamically preferred. At \(\mu=\mu_c\), the large and small black hole phases merge and the first-order transition vanishes. Thus \(\mu_c\) marks a second-order critical point where the first-order line terminates. This behavior qualitatively matches expectations from lattice QCD at finite density and temperature\cite{deForcrand:2002hgr,Stephanov:2004wx,Borsanyi:2025dyp}. The small-large black hole transition has been linked to the confinement-deconfinement transition in Ref.~\cite{He:2013qq}. The authors of Ref.~\cite{Dudal:2017max} treats it as a specious-confinement-deconfinement transition. Exact confinement requires a vanishing Polyakov loop. In the small black hole phase, the Polyakov loop is nonzero, although extremely small. At low \(T\), it still exhibits linear confinement at large distances. This is why the phase is called the specious-confinement phase.

\section{Spectral function from the self-consistent holographic EMD model}\label{sec:03}

We adopt the self-consistent EMD background and study the spectral functions of heavy and exotic
mesons. We compute the spectral functions using the membrane paradigm of Ref.~\cite{Iqbal:2008by} and the spectral function is evaluated following Ref.~\cite{Braga:2018zlu}. The two frameworks use different but equivalent probe actions. The corresponding action is given by \cite{Braga:2018zlu}:
\begin{equation}\label{eq:x3}
I= \int d^{4}x dz \sqrt{-g} e^{-\phi_s(z)}\bigg[-\frac{1}{4g^2_5 }F_{mn}F^{mn} \bigg],
\end{equation}
where $F_{mn}=\partial_m V_n -\partial_n V_m$. For simplicity, the coupling $g^2_5$ is set to unity.

Comparing the action of the  EMD background (Eq.(\ref{eqa112})), we identify $e^{\phi_s(z)}=\frac{1}{f(z)}=e^{A(z)+B(z)}.$ $B(z)=(\kappa z)^{2-\alpha} + Mz + \tanh\left(\frac{1}{Mz} -\frac{\kappa}{\sqrt{\Gamma}}\right)\Big]$. The values of $B(z)$ parameters for the four meson states are in the Table1. Therefore, the membrane-paradigm formulas can be applied directly to the self-consistent background  by replacing $e^{\phi_s(z)}$ with \(1/f(z)\). The background itself remains the self-consistent EMD solution, so the backreaction on the metric is fully retained. The backreaction of the self-consistent background is fully included in the spectral function calculation. In the EMD framework, the dilaton, gauge field, and metric are solved simultaneously. Therefore, the function \(f(z)\) that enters the probe sector already carries the imprint of the medium on the geometry. When we compute the spectral function, this imprint enters the probe dynamics through the background. At finite chemical potential, the charge density modifies the blackening function \(g(z)\) through terms involving \(f(z)\). So the calculation is not a probe on a fixed background, but a probe on a background that has already responded to the medium.

Following the membrane paradigm, we derive the heavy-meson spectral function. To proceed, we rewrite the background metric, Eq.~\eqref{eq18}, in a general form
\begin{equation}
\label{eqa6}
 ds^2= -g_{tt}dt^2 +g_{xx_{1}} dx_{1}^{2}+ g_{xx_{2}} dx_{2}^{2}+g_{xx_{3}} dx_{3}^{2}+g_{zz}dz^2,
\end{equation}
where $g_{tt}=\frac{L^2 e^{2 A(z)}}{z^2} g(z)$, $g_{xx_{1}}=g_{xx_{2}}=g_{xx_{3}}=\frac{L^2 e^{2 A(z)}}{z^2}$, and $g_{zz}=\frac{L^2 e^{2 A(z)}}{g(z) z^2}$.

Varying Eq.~\eqref{eq:x3} yields the equation of motion
\begin{equation}
\label{eqa8}
 \partial^m \bigg(\frac{\sqrt{-g}}{h(z)}F_{mn}\bigg)=0,
\end{equation}
where we set $h(z)=e^{\phi_s(z)}=\frac{1}{f(z)}$.

For a \(z\)-foliation, the conjugate momentum of the gauge field is given by
\begin{equation}
\label{eqa9}
 j^\mu= -Q F^{z\mu}.
\end{equation}

Equation~\eqref{eq18} describes a background that is isotropic in the \(\vec{x}\)-direction. The equations of motion split into longitudinal fluctuations in the \((t,x_3)\) plane and transverse fluctuations in the \((x_1,x_2)\) plane. For Eq.~\eqref{eqa8}, the longitudinal components are
\begin{equation}
\label{eqav8}
\begin{split}
&  -\partial_z j^t -\frac{\sqrt{-g}}{h(z)}g^{tt}g^{xx_3}\partial_{x_3}F_{x_3 t}=0,\\
 & -\partial_z j^{x_3} +\frac{\sqrt{-g}}{h(z)}g^{tt}g^{xx_3}\partial_{t}F_{x_3 t}=0.\\
 & \partial_{x_3} j^{x_3}+\partial_t j^t =0.
 \end{split}
\end{equation}

Using the Bianchi identity, we obtain
\begin{equation}
\label{eqa9}
 \partial_z F_{x_3 t}-\frac{h(z)}{\sqrt{-g}}g_{zz}g_{xx_3}\partial_t j^z -\frac{h(z)}{\sqrt{-g}} g_{tt}g_{xx_3}\partial_{x_3}j^t=0.
\end{equation}

For the longitudinal channel, the conductivity and its derivative are
\begin{equation}
\label{eqav9}
\begin{split}
&  \overline{\sigma}_L (\omega,\overrightarrow{p},z)=\frac{j^{x_3}(\omega,\overrightarrow{p},z)}{F_{x_3 t}(\omega,\overrightarrow{p},z)},\\
 & \partial_z \overline{\sigma}_L=-i\omega\sqrt{\frac{g_{zz}}{g_{tt}}}\bigg[\Sigma(z)-\frac{\overline{\sigma}^2_L}{\Sigma(z)}(1-\frac{p^2_3}{\omega^2}\frac{g^{xx_3}}{g^{tt}}) \bigg],
 \end{split}
\end{equation}
with \(p=(\omega,0,0,p_3)\) and
\(\Sigma(z)=\frac{1}{h(z)}\sqrt{\frac{-g}{g_{zz}g_{tt}}}\,g^{x_3 x_3}\).

The transverse channel yields
\begin{equation}
\label{eqa10}
 \partial_z \overline{\sigma}_T=i\omega \sqrt{\frac{g_{zz}}{g_{tt}}} \bigg[\frac{\overline{\sigma}^2_T}{\Sigma(z)}- \Sigma(z)(1-\frac{p^2_3}{\omega^2}\frac{g^{xx_3}}{g^{tt}})  \bigg].
\end{equation}

For \(p_3=0\), the conductivities become equal, \(\bar{\sigma}=\bar{\sigma}_T=\bar{\sigma}_L\). The explicit form is
\begin{equation}
\label{eqa11}
 \partial_z \overline{\sigma}=i\omega \sqrt{\frac{g_{zz}}{g_{tt}}} \bigg[\frac{\overline{\sigma}^2}{\Sigma(z)}- \Sigma(z)  \bigg].
\end{equation}

From the Kubo formula, the AC conductivity \(\sigma\) reads
\begin{equation}
\label{eqa12}
 \sigma(\omega)=-\frac{G_R (\omega)}{i \omega}\equiv \overline{\sigma}(\omega,z=0).
\end{equation}

We then obtain the spectral function
\begin{equation}
\label{eqa131}
 \rho(\omega)\equiv -Im G_R (\omega)=\omega Re \overline{\sigma}(\omega,0).
\end{equation}

\begin{figure}[H]
    \centering
      \setlength{\abovecaptionskip}{0.1cm}
    \includegraphics[width=10cm]{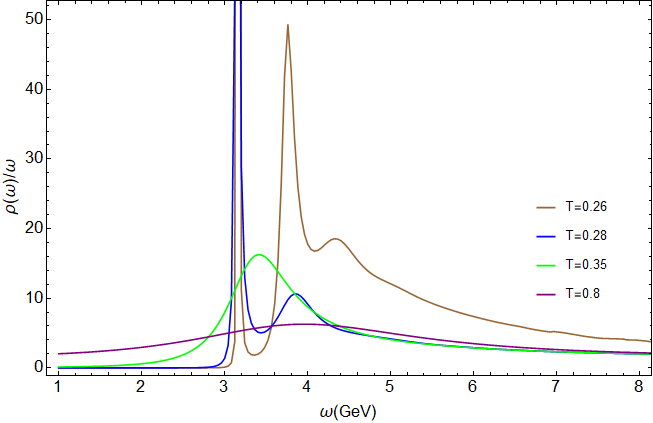}
    \caption{\label{fig4} Spectral functions of charmonium at $\mu=0.1$ GeV for different $T$.}
\end{figure}

\begin{figure}[H]
    \centering
      \setlength{\abovecaptionskip}{0.1cm}
    \includegraphics[width=10cm]{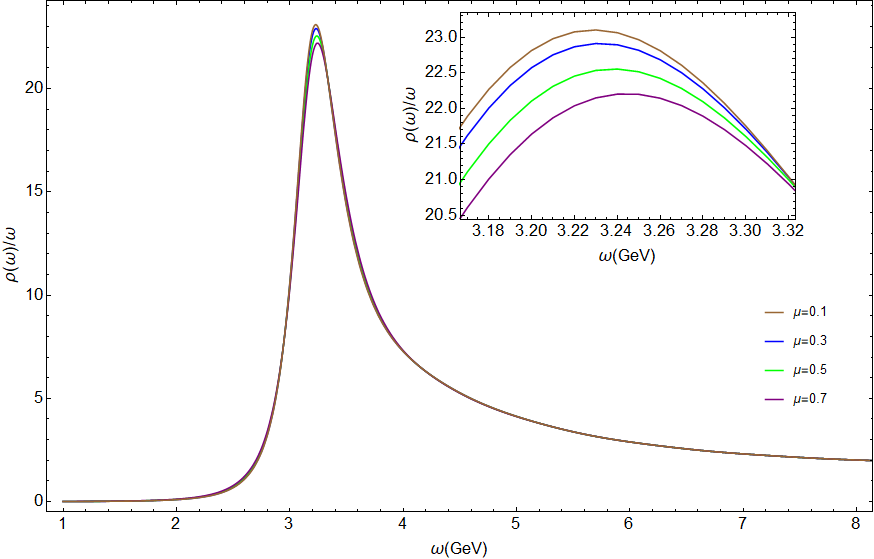}
    \caption{\label{fig5} Spectral functions of charmonium at $T=0.3$ GeV for different $\mu$.}
\end{figure}

We define the melting temperature \(T_{\rm melt}\) as the temperature at which the spectral peak no longer exhibits a local maximum. This criterion is applied uniformly to all four states. In Fig.~\ref{fig4}, we display the charmonium spectral functions at \(\mu=0.1\) for various temperatures. The first, second, and third peaks in the bell-shaped profiles correspond to the 1S (\(J/\psi\)), 2S (\(\psi(2S)\)), and 3S (\(\psi(4040)\)) states. Raising \(T\) causes all peaks to broaden and shrink, indicating that charmonium melts progressively. The 3S state vanishes completely near \(T=T_c\), with \(T_c=0.27\) GeV, and the 2S state dissolves at roughly \(T_{\rm melt}=1.07T_c\). The 1S state disappears at \(T=0.8\) GeV, corresponding to \(T_{\rm melt}=2.96T_c\). Lattice QCD simulations suggest that \(J/\psi\) dissociates gradually and disappears near \(3T_c\) \cite{Datta:2003ww}, which agrees well with our findings. In Fig.~\ref{fig5}, we plot the \(J/\psi\) spectral function at fixed temperature for several chemical potentials. As \(\mu\) grows, the peak becomes lower and broader, which signals stronger dissociation of \(J/\psi\) in a denser medium.

\begin{figure}[H]
    \centering
      \setlength{\abovecaptionskip}{0.1cm}
    \includegraphics[width=10cm]{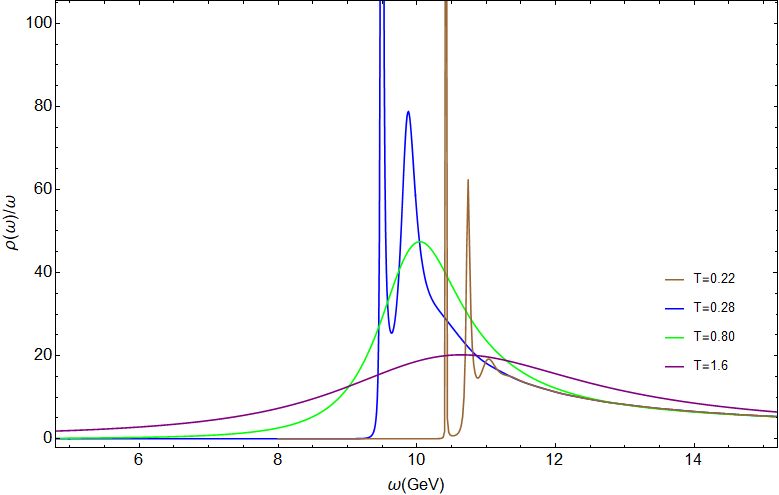}
    \caption{\label{fig6} Spectral functions of bottomonium at $\mu=0.1$ GeV for different $T$.}
\end{figure}

\begin{figure}[H]
    \centering
      \setlength{\abovecaptionskip}{0.1cm}
    \includegraphics[width=10cm]{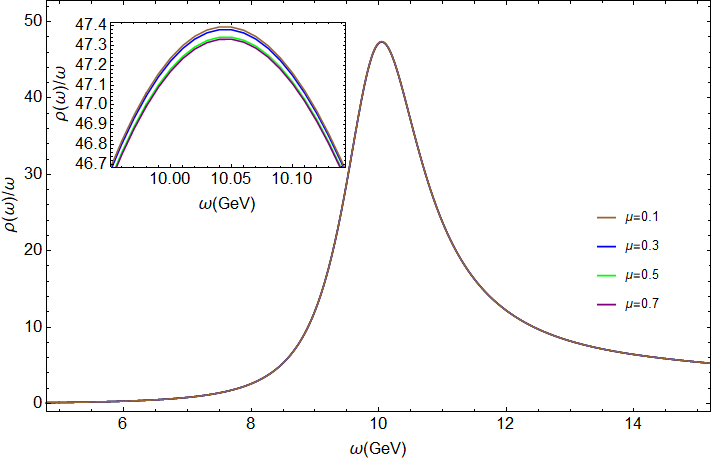}
    \caption{\label{fig7} Spectral functions of bottomonium at $T=0.8$ GeV for different $\mu$.}
\end{figure}

Fig.~\ref{fig6} presents the bottomonium spectral functions at \(\mu=0.1\) for several temperatures. Three peaks appear in the spectral profiles, corresponding to the 1S (\(\Upsilon(1S)\)), 2S (\(\Upsilon(2S)\)), and 3S (\(\Upsilon(3S)\)) states, respectively. As the temperature rises, all peaks become broader and lower, which signals the progressive melting of bottomonium in the hot medium. The \(\Upsilon(1S)\) state disappears at \(T_{\rm melt}=1.6\) GeV, i.e. \(T_{\rm melt}=5.93T_c\). The results show that the excited states dissolve before the ground state. Moreover, bottomonium survives to higher temperatures than charmonium, because the larger bottom quark mass leads to stronger binding. CMS and STAR measurements show that the \(\Upsilon(2S)\) state dissociates at a lower temperature than the \(\Upsilon(1S)\) state \cite{CMS:2018zza,STAR:2022rpk}. Lattice QCD calculations indicate that charmonia have a lower dissociation temperature than bottomonia, so they melt more easily in the QGP due to weaker binding \cite{Burnier:2015tda}. Our results agree well with these simulations.

In Fig.~\ref{fig7}, the \(\Upsilon(1S)\) spectral function is plotted at fixed temperature for different chemical potentials. With increasing \(\mu\), the peak drops and broadens, indicating enhanced dissociation in a dense medium. This trend is similar to the \(J/\psi\) case, where a larger \(\mu\) also accelerates melting.

\begin{figure}[H]
    \centering
      \setlength{\abovecaptionskip}{0.1cm}
    \includegraphics[width=10cm]{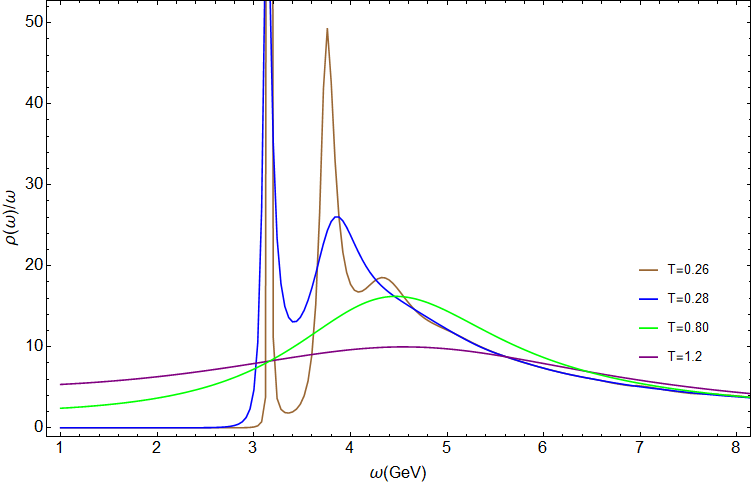}
    \caption{\label{fig8} Spectral functions of tetraquark at $\mu=0.1$ GeV for different $T$.}
\end{figure}

\begin{figure}[H]
    \centering
      \setlength{\abovecaptionskip}{0.1cm}
    \includegraphics[width=10cm]{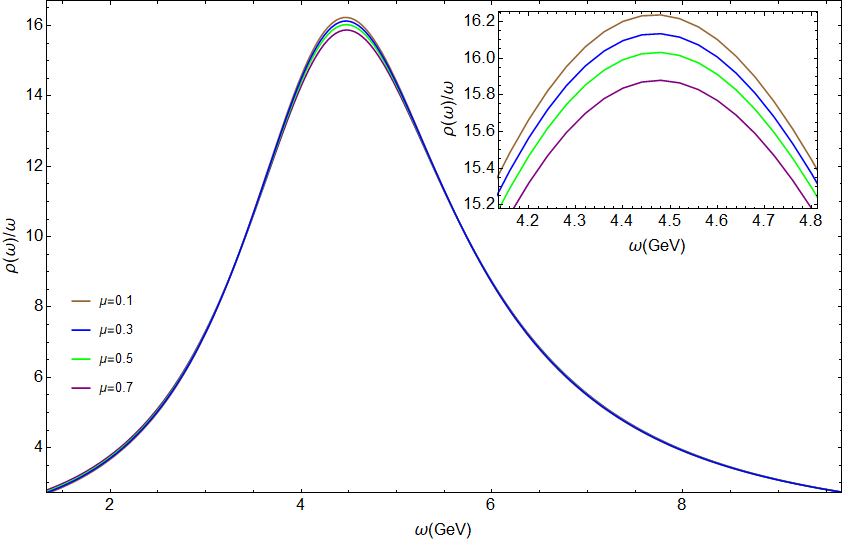}
    \caption{\label{fig9} Spectral functions of tetraquark at $T=0.8$ GeV for different $\mu$.}
\end{figure}

Fig.~\ref{fig8} shows the tetraquark spectral functions at \(\mu=0.1\) for several temperatures. Three peaks appear in the bell-shaped profiles, corresponding to the 1S (\(Z_c(3900)\)), 2S (\(Z_c(4200)\)), and 3S (\(Z_c(4430)\)) states, respectively. As the temperature rises, all peaks broaden and decrease in height, which indicates that the tetraquark states melt progressively in the hot medium. The 1S state disappears at \(T=1.2\) GeV, corresponding to \(T_{\rm melt}=4.44T_c\). Fig.~\ref{fig9} displays the \(Z_c(3900)\) spectral function at fixed temperature for several chemical potentials. With increasing \(\mu\), the peak becomes lower and broader, which points to stronger dissociation of \(Z_c(3900)\) in a denser medium.

\begin{figure}[H]
    \centering
      \setlength{\abovecaptionskip}{0.1cm}
    \includegraphics[width=10cm]{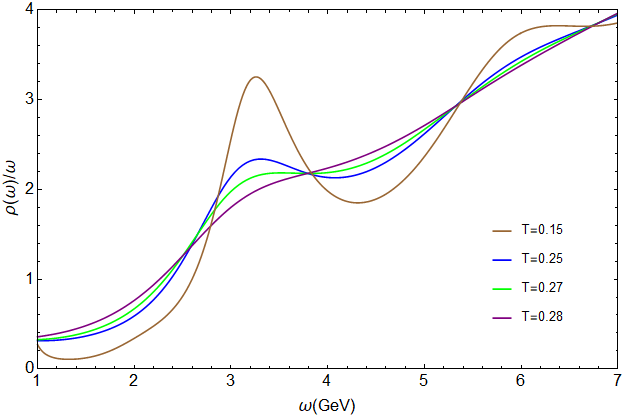}
    \caption{\label{fig10} Spectral functions of hybrid meson at $\mu=0.1$ GeV for different $T$.}
\end{figure}

Fig.~\ref{fig10} presents the hybrid meson spectral functions at \(\mu=0.1\) for different temperatures. The two peaks correspond to the 1S (\(\pi_1(1400)\)) and 2S (\(\pi_1(1600)\)) states. As \(T\) increases, both peaks broaden and drop in height, showing the progressive melting of the hybrid meson. The \(\pi_1(1400)\) state vanishes at \(T_{\rm melt}=0.28\) GeV (\(T_{\rm melt}=1.08T_c\)), much lower than the melting temperatures of conventional quarkonia and tetraquarks.

\begin{figure}[H]
    \centering
      \setlength{\abovecaptionskip}{0.1cm}
    \includegraphics[width=10cm]{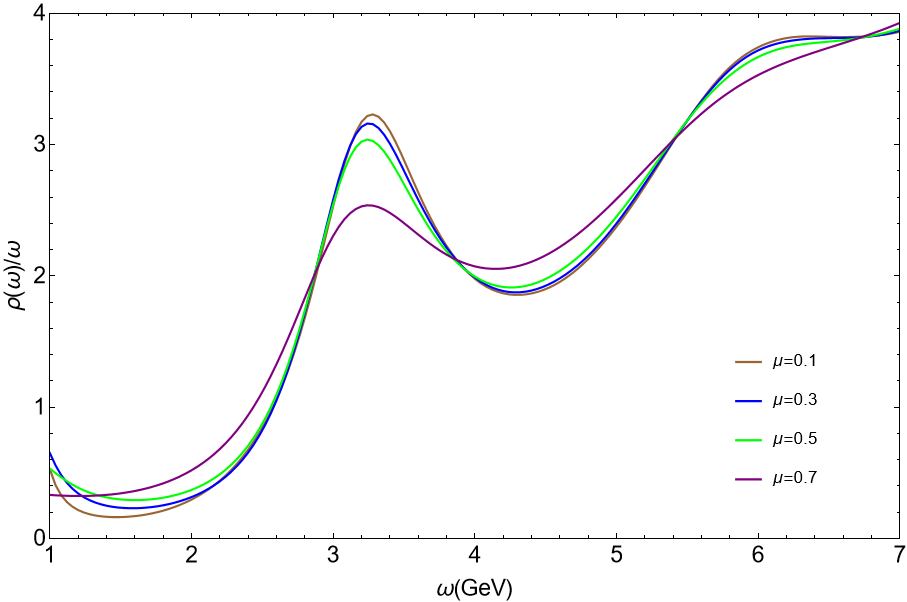}
    \caption{\label{fig11} Spectral functions of hybrid meson at $T=0.15$ GeV for different $\mu$.}
\end{figure}

In Fig.~\ref{fig11}, the hybrid meson spectral function is plotted at fixed temperature for several chemical potentials. A larger \(\mu\) lowers and broadens the peak, indicating stronger dissociation in a dense medium. This behavior resembles that of \(J/\psi\) and other heavy states, though the hybrid meson is the most easily dissociated.

Our work extends the self-consistent EMD model and compute spectral functions for all four heavy and exotic mesons in both the 1S and 2S channels. The soft-wall model of Ref.~\cite{Braga:2017bml} differs in an important way. There the dilaton does not backreact on the metric, so the background stays fixed and medium effects enter only through the probe. Here the dilaton is part of the self-consistent EMD solution. Thus the spectral functions are tied to the QCD thermodynamics of the background rather than to a phenomenological dilaton profile, thus our predictions are more constrained. We further find that the effective mass of heavy quarkonium grows with temperature and chemical potential, consistent with Ref.~\cite{Zhao:2021ogc}.

\section{Conclusion and discussion}\label{sec:04}

In this work, we have investigated the spectral functions of heavy and exotic mesons
within a self-consistent holographic Einstein-Maxwell-Dilaton (EMD) background at finite temperature and chemical potential. By solving the coupled Einstein, Maxwell, and dilaton equations simultaneously, with the dilaton back-reacting on the metric, we obtain a holographic dual of QCD that captures key non-perturbative features, including confinement, deconfinement, and meson melting. This self-consistent treatment distinguishes our approach from earlier bottom-up holographic studies in which the dilaton profile was imposed by hand rather than determined dynamically.

Our central finding is that the dissolution temperatures of the $1S$ states follow
the same ordering as their masses: $T_{\text{melt}}^{\pi_1(1400)} < T_{\text{melt}}^{J/\psi}
< T_{\text{melt}}^{Z_c(3900)} < T_{\text{melt}}^{\Upsilon(1S)}.$ This monotonic relation between mass and melting temperature is consistent with the physical intuition that a larger quark mass produces stronger binding, so that more thermal energy is required to break the bound state. It also aligns qualitatively
with the sequential-melting picture established at RHIC and the LHC, where the more loosely bound excited states dissociate at lower temperatures than the tightly bound ground states. The fact that the $2S$ state melts before the $1S$ state in every channel we examined lends further support to this interpretation and is consistent with the well-documented sequential suppression of quarkonium in heavy-ion collisions.

Beyond the expected $1S$/$2S$ ordering, our results reveal a clear hierarchy among the different meson species. The hybrid meson $\pi_1(1400)$ is the most fragile state in our analysis, dissolving at $T \simeq 0.28$ GeV ($1.08\,T_c$), while $\Upsilon(1S)$ survives up to $T \simeq 1.6$ GeV ($5.93\,T_c$). The intermediate states, $J/\psi$ at $0.8$ GeV ($2.96\,T_c$) and $Z_c(3900)$ at $1.2$ GeV ($4.44\,T_c$), fall neatly between these extremes. This ordering is notable because the internal structure of $Z_c(3900)$ remains under debate; our results suggest that, whatever its precise nature, its binding is intermediate between charmonium and bottomonium, with a correspondingly intermediate melting temperature. This
provides a nontrivial consistency check on the model, which was not tuned to reproduce this specific ordering.

Our results for $Z_c(3900)$ and $\pi_1(1400)$ can be regarded as predictions for future experimental searches at ALICE and CMS. Because both states are expected to melt at relatively low temperatures, with \(4.44\,T_c\) for \(Z_c(3900)\) and \(1.08\,T_c\) for \(\pi_1(1400)\), their yields in heavy-ion collisions should be strongly suppressed relative to the expectation from a superposition of nucleon-nucleon collisions. A measurement of their nuclear modification factors, and in particular a comparison of their suppression patterns with those of $J/\psi$ and $\Upsilon(1S)$, would provide a direct test of the hierarchy predicted here. We caution that connecting the spectral function at finite \(T\) and \(\mu\) to experimentally observed yields is not straightforward, since effects such as regeneration, feed-down from excited states, and final-state interactions lie beyond the present scope. The melting temperatures quoted here should therefore be interpreted as thresholds for the disappearance of the quasiparticle peak, not as direct predictions for observable suppression factors.

Several limitations of this study should be acknowledged. First, our analysis is restricted to vanishing magnetic field and zero angular velocity, whereas heavy-ion collisions generate strong magnetic fields and substantial vorticity in their early stages, which can induce anisotropic dissociation of different spin states. The melting temperatures reported here should thus be regarded as baseline values that will be modified in more realistic settings. Second, although the EMD background captures confinement and deconfinement, it does not include dynamical quarks in the fundamental representation; the mesons are treated as probes in a fixed gravitational background. This is a common approximation in holographic QCD,
but it neglects the possible back-reaction of the mesonic degrees of freedom on the gluonic sector. Third, our determination of the melting temperature relies on the disappearance of a discernible peak in the spectral function, which inevitably involves a degree of judgment; a more rigorous criterion, such as the disappearance of the quasiparticle pole, would be desirable.

These limitations point to natural directions for future work. The most immediate extension is to include a finite magnetic field or rotation, which would permit the study of anisotropic dissociation and spin-dependent melting patterns that are inaccessible in the present isotropic setup. It would also be interesting to incorporate finite quark-mass effects or the back-reaction of the meson fields, bringing the model closer to a fully self-consistent treatment. Finally, a detailed comparison with data from ALICE and CMS, in particular the nuclear modification factors of \(J/\psi\), \(\Upsilon\), and, if measurable, \(Z_c\), would help constrain the model and identify which aspects of the holographic description are robust and which require refinement.

In summary, we have shown that a self-consistent holographic EMD background reproduces the expected sequential melting of heavy and exotic mesons and predicts a clear hierarchy of dissolution temperatures that correlates with particle mass: the hybrid meson $\pi_1(1400)$ emerges as the most fragile state, $\Upsilon(1S)$ as the most robust, and $J/\psi$ and $Z_c(3900)$ in between. These results provide
experimentally testable predictions and motivate further work incorporating magnetic fields, rotation, and a more complete treatment of the quark sector.

\section*{Acknowledgments}

Zhou-Run Zhu is supported by the National Natural Science Foundation of China under Grant No.12605236. Defu Hou is supported in part by the National Key Research and Development Program of China under Contract No.2022YFA1604900. Defu Hou is also supported by the National Natural Science Foundation of China(NSFC) under Grants No.12435009,and No.12275104.

\section*{Declarations}

\textbf{Conflict of interest}  The authors declare that they have no Conflict of interest.

\end{document}